\documentclass[prb,twocolumn,superscriptaddress,a4paper,twoside]{revtex4}

\usepackage{amsmath}
\usepackage{amssymb}
\usepackage{todonotes}
\usepackage{nicefrac}
\usepackage{graphics}
\usepackage{color}
\usepackage{graphicx}
\usepackage{verbatim}
\usepackage{float}

\makeatletter
\newcommand*{\rom}
[1]{\expandafter\@slowromancap\romannumeral #1@}
\makeatother

\begin{document}

\title{Continuity equation for the many-electron spectral function}

\author{F. Aryasetiawan}
\affiliation{
Department of Physics, Division of Mathematical Physics, 
Lund University, Professorsgatan 1, 223 63, Lund, Sweden}
\affiliation{LINXS Institute of advanced Neutron and X-ray Science (LINXS), IDEON Building: Delta 5, Scheelevägen 19, 223 70 Lund, Sweden}

\begin{abstract}
Starting
from the recently proposed dynamical exchange-correlation field framework, 
the equation of motion of the diagonal part of the many-electron Green function is derived, from which
the spectral function can be obtained. The resulting equation of motion takes the form of 
the continuity equation of charge and current densities in electrodynamics with a source. 
An unknown quantity in this equation is the current density, corresponding to the kinetic energy.
A procedure à la Kohn-Sham scheme is then proposed, 
in which the difference between the kinetic potential of the
interacting system and the non-interacting Kohn-Sham system is shifted into the exchange-correlation field.
The task of finding a good approximation for the exchange-correlation field should be
greatly simplified since
only the diagonal part is needed. A formal
solution to the continuity equation provides an explicit expression
for calculating the spectral function, given an approximate exchange-correlation field.
\end{abstract}

\maketitle

\section{Introduction}
The total spectral function of a many-electron system, hereafter referred to simply as 
the spectral function, is given by the trace of
the Green function. This implies that to calculate the spectral function
only the diagonal components of the
Green function are required. Although for solids, the momentum-resolved
spectral function contains more detailed information about the electronic 
structure of the system, 
it often suffices for many purposes to know the integrated spectral
function. It is therefore an attractive proposition to determine the spectral 
function from the diagonal part of the Green function since it is presumably 
much simpler to calculate than the full Green function.
A relevant work along this direction is the work by
Gatti \emph{et al} \cite{gatti2007} who proposed using an effective potential, local in space
but energy dependent, from which
the spectral function can be calculated directly. It is quite feasible that for 
a given system an effective potential
that reproduces the exact diagonal part of the Green function exists. 
It is, however, not evident how to construct such an effective potential.
Another work of relevance is that of Savrasov and Kotliar \cite{savrasov2004}, who introduced the
concept of spectral density-functional theory. In their work, the key variable is given
by the local Green function rather than the electron density. 

In this paper a different approach is taken. Starting from a recently
derived equation of motion of the Green function within
the dynamical exchange-correlation field framework \cite{aryasetiawan2022a,aryasetiawan2022b}, 
an equation of motion for  the diagonal
part of the Green function, is obtained. The derivation takes
advantage of the fact that the exchange-correlation field acts locally
on the Green function. It should be noted that 
a similar derivation cannot be followed in a natural way within the
self-energy formalism. The resulting equation has the
form of the continuity equation of charge and current densities 
in electrodynamics with a source/sink term.
An unknown quantity in the equation is the current density, which can be associated
with the kinetic energy.
By introducing the Kohn-Sham current density and transferring the difference in
kinetic energy between
the interacting system and the non-interacting Kohn-Sham system into
the exchange-correlation field, a formally
exact continuity equation for the diagonal part of the Green function is 
obtained. For practical calculations, a local-density approximation for the
modified exchange-correlation field based
on the homogeneous electron gas is proposed.
An example from a model of the interacting electron gas is considered to
illustrate the exchange-correlation field and the kinetic potential.

The paper continues with a theory section, deriving the continuity equation,
followed by an illustration from the model electron gas. It closes with
a summary and conclusions.

\section{Theory}

The equation of motion of the Green function in the dynamical
exchange-correlation (xc) field framework
is given by \cite{aryasetiawan2022a}
\begin{equation}
\left(  i\partial_t-h(r\mathbf{)}-V_\mathrm{xc}(r,r^{\prime
};t)\right)  G(r,r^{\prime};t)=\delta(r-r^{\prime})\delta(t), 
\label{EOM}%
\end{equation}
where 
\begin{equation}
    h(r)=-\frac{1}{2}\nabla^2 +V_\mathrm{MF}(r), \quad 
    V_\mathrm{MF}=V_\mathrm{ext}+V_\mathrm{H}.
\end{equation}
$r$ is a combined label for position and spin: $r=(\mathbf{r},\sigma)$ and
$\int dr = \sum_\sigma\int d^3r$.

A temporal density proportional to the diagonal part of the Green function
is defined as follows:
\begin{align}
    \rho(r,t)&=-iG(r,r;t).
\end{align}
For $t=0^-$ the temporal density reduces to the electron density:
\begin{align}
    \rho(r,0^-)&=-iG(r,r;0^-)=\rho(r).
\end{align}
When $\rho(r,t)$ is
integrated over $r$ and Fourier transformed in $t$, it yields the spectral function
or density of states:
\begin{equation}
    \rho(\omega)=\frac{1}{\Omega}\int dr\int dt\, e^{i\omega t}\rho(r,t).
\end{equation}

Considering the equation of motion for the Green function
in Eq. (\ref{EOM}) for $r'=r$ and defining
\begin{align}
    V_\mathrm{xc}(r,t)=V_\mathrm{xc}(r,r;t),
\end{align}
one finds
\begin{align}
    &\left[ i\partial_t -V_\mathrm{MF}(r)
    -V_\mathrm{xc}(r,t) \right]\rho(r,t)
    \nonumber\\
    &\qquad 
    -\left.\frac{i}{2} \nabla^2 G(r,r';t)\right|_{r'=r}=0.
\end{align}
By defining a current density
\begin{align}
    \mathbf{j}(r,t)=-\frac{1}{2}\left.\nabla G(r,r';t)\right|_{r'=r},
\end{align}
the equation of motion for $t\neq 0$ becomes
\begin{align}
    \partial_t\rho(r,t) + \nabla \cdot \mathbf{j}(r,t)=S(r,t),
    \label{eq:continuity}
\end{align}
where
\begin{align}
    S(r,t)= -i\left[V_\mathrm{MF}(r) +V_\mathrm{xc}(r,t)\right]\rho(r,t).
\end{align}
This can be interpreted as a continuity equation with a source/sink
term $S$ on the right-hand side. Since the divergence of
the current density is the curvature
of the Green function at $r'=r$, only knowledge of the diagonal components, 
$G(r,r;t)$,
and the neighboring points along the diagonal, $G(r\pm \delta r,r;t)$,
is needed. Substantially much less information than that of the full Green function
is required to calculate the spectral function.
There is no auxiliary system invoked in this derivation 
and all quantities are well defined and their existence are guaranteed.

Integrating the continuity equation in space
yields
\begin{align}
    \partial_t\rho(t)+\int_{A(V)} d\mathbf{A} \cdot \mathbf{j}(r,t) =\int dr \,S(r,t),
\end{align}
where \begin{align}
    \rho(t) = \int dr \,\rho(r,t).
\end{align}
Gauss' theorem has been used:
\begin{align}
    \int_V dr \nabla \cdot \mathbf{j}(r,t) = \int_{A(V)} d\mathbf{A} \cdot \mathbf{j}(r,t).
\end{align}

The continuity equation can be rewritten as follows:
\begin{align}
   i \partial_t \ln{\rho(r,t)} = V_\mathrm{MF}(r) +V_\mathrm{xc}(r,t)+V_\mathrm{K}(r,t),
\end{align}
where $V_\mathrm{K}$ is the kinetic potential,
\begin{align}
    V_\mathrm{K}(r,t)= -i\frac{\nabla \cdot \mathbf{j}(r,t)}{\rho(r,t)}.
\end{align}
The formal solution is given by
\begin{align}
    \rho(r,t)&=\rho(r) \exp{\left[-iV_\mathrm{MF}(r) t\right]}
    \nonumber\\
    &\times\exp{\left\{-i\int_0^t dt'
    \left[ {V}_\mathrm{xc}(r,t') +V_\mathrm{K}(r,t')
    \right] \right\}}.
    \label{eq:formalsoln}
\end{align}
Alternatively,
\begin{align}
    \rho(r,t)=\rho(r)+\int_0^t dt'\left[ S(r,t')-\nabla \cdot \mathbf{j}(r,t')\right].
\end{align}
Assuming that a good approximation for $V_\mathrm{xc}$ is known, the remaining
input required to solve
for the temporal density is the current density $\mathbf{j}$. The current density 
is associated with the kinetic energy, which is known to be very difficult to
approximate with an explicit functional of the electron density.

To construct a practical scheme for calculating the temporal density,
one may follow the Kohn-Sham scheme of density functional theory 
\cite{kohn1965,jones1989,becke2014,jones2015}
by defining $\Delta V_\mathrm{K}$ according to
\begin{align}
\Delta V_\mathrm{K} = -i\left\{ \frac{\nabla \cdot \mathbf{j}}{\rho} 
-\frac{\nabla \cdot \mathbf{j}^\mathrm{KS}}{\rho^\mathrm{KS}}\right\}
= V_\mathrm{K}-V_\mathrm{K}^\mathrm{KS},
\end{align}
where $\rho^\mathrm{KS}$ and $\mathbf{j}^\mathrm{KS}$ are the temporal density and the current density
obtained from
the Kohn-Sham Green function. 
$\Delta V_\mathrm{K}$ may be interpreted as the difference in kinetic potential between the
interacting system and the non-interacting Kohn-Sham system.
The continuity equation becomes
\begin{align}
    \partial_t\rho(r,t) + \frac{\rho(r,t)}{\rho^\mathrm{KS}(r,t)}\nabla \cdot \mathbf{j}^\mathrm{KS}(r,t)=\Tilde{S}(r,t),
    \label{eq:continuity1}
\end{align}
or
\begin{align}
   i \partial_t \ln{\rho(r,t)} = V_\mathrm{MF}(r) +\Tilde{V}_\mathrm{xc}(r,t)
    +V^\mathrm{KS}_\mathrm{K}(r,t),
\end{align}
where
\begin{align}
    \Tilde{S}(r,t)= -i\left[V_\mathrm{MF}(r) +\Tilde{V}_\mathrm{xc}(r,t)\right]\rho(r,t),
\end{align}
\begin{align}
    \Tilde{V}_\mathrm{xc} = {V}_\mathrm{xc} + \Delta V_\mathrm{K}.
\end{align}

The formal solution is given by
\begin{align}
    \rho(r,t)&=\rho(r) \exp{\left[-iV_\mathrm{MF}(r) t\right]}
    \nonumber\\
    &\times\exp{\left\{-i\int_0^t dt'
    \left[ \Tilde{V}_\mathrm{xc}(r,t') + V_\mathrm{K}^\mathrm{KS}(r,t')
    \right] \right\}},
\end{align}
or alternatively,
\begin{align}
    \rho(r,t)=\rho(r)+\int_0^t dt'\left[ \Tilde{S}(r,t')-\frac{\rho(r,t)'}{\rho^\mathrm{KS}(r,t')}\nabla \cdot 
    \mathbf{j}^\mathrm{KS}(r,t')\right].
\end{align}

This procedure is analogous to the Kohn-Sham scheme \cite{kohn1965,jones1989,becke2014,jones2015}
in which the difference in kinetic
energy between the interacting system and the auxiliary non-interacting system
is shifted into
the exchange-correlation energy. Here, the difference between
the kinetic potentials of the
interacting system and the non-interacting Kohn-Sham system is incorporated into the
exchange-correlation field.
The problem of calculating the spectral function amounts to finding a good approximation for
$\Tilde{V}_\mathrm{xc}$, which should be much simpler compared with the 
full exchange-correlation field that depends on two position variables.
$\Tilde{V}_\mathrm{xc}(n,t)$ can be calculated for the homogeneous electron gas (HEG) as
a function of the electron density $n$ within, e.g., the $GW$ approximation 
\cite{hedin1965,hedin1969,aryasetiawan1998} or better 
approximations such as the cumulant 
expansion \cite{langreth1970,bergersen1973,hedin1980,almbladh1983,aryasetiawan1996,kas2014}, and applied to
real inhomogeneous systems within, for example, the local-density approximation (LDA):
\begin{align}
    \Tilde{V}_\mathrm{xc}^\mathrm{LDA}(r,t)=\Tilde{V}_\mathrm{xc}^\mathrm{HEG}(\rho(r),t).
\end{align}

\subsection{Non-interacting homogeneous electron gas}

As an example, consider
the non-interacting homogeneous electron gas whose Green function is given by:
\begin{align}
    iG_0(R;t) &= \frac{1}{\Omega} \sum_{k> k_\mathrm{F}} 
    e^{i\mathbf{k}\cdot \mathbf{R}} e^{-i\varepsilon_k t}\theta(t)
    \nonumber\\
    &-\frac{1}{\Omega} \sum_{k\leq k_\mathrm{F}} 
    e^{i\mathbf{k}\cdot\mathbf{R}-\mathbf{r}'} e^{-i\varepsilon_k t}\theta(-t),
    \label{eq:G0}
\end{align}
where $\varepsilon_k=\frac{1}{2}k^2$, $k_\mathrm{F}$ 
is the Fermi wave vector, and $\Omega$ is
the space volume. 

For a non-interacting electron gas $V_\mathrm{xc}=0$ and
$V_\mathrm{ext}$ is a uniform positive background
so that $V_\mathrm{MF}=V_\mathrm{ext}+V_\mathrm{H}=0$. Since the system is uniform,
only the case of $R=0$ is needed. 
The temporal density per spin is given by
\begin{align}
    \rho_0(t<0)&=
    \frac{1}{\Omega} \sum_{k\leq k_\mathrm{F}} 
    e^{-i\varepsilon_k t}
    \nonumber\\
    &=\frac{1}{2\pi^2} \int_0^{k_\mathrm{F}} dk\,k^2e^{-itk^2/2}
    \label{eq:rho0}
\end{align}
and the kinetic energy corresponding to the current density is given by
\begin{align}
    \nabla \cdot \mathbf{j}_0(r,t)&=-\left.\frac{1}{2} \nabla^2 G_0(R;t<0)\right|_{R=0}
    \nonumber\\
    &=\frac{i}{\Omega} \sum_{k\leq k_\mathrm{F}} 
    \frac{k^2}{2} e^{-i\varepsilon_k t}
    \nonumber\\
    &=\frac{i}{4\pi^2} \int_0^{k_\mathrm{F}} dk\,k^4e^{-itk^2/2},
\end{align}
where
\begin{equation}
    k_\mathrm{F}^3 = 3\pi^2 n_0,\quad \rho_0(0)=\frac{1}{2}n_0,
\end{equation}
and $n_0$ is the density of the homogeneous electron gas. Since for 
the non-interacting electron gas $S=0$, the continuity equation in (\ref{eq:continuity})
is indeed fulfilled.

\subsection{A model Green function for the interacting electron gas}

To illustrate and study the behavior of the exchange-correlation field and the kinetic potential, 
a physically motivated model for the Green function of the interacting electron gas
is considered. This model was proposed in a previous article \cite{karlsson2023}
and given by the following:
\begin{align}
    &G(R,t<0) =\frac{i}{\Omega}\sum_{k\leq k_\mathrm{F}} (C_1 +C_2 )
     e^{i\mathbf{k}\cdot\mathbf{R}},
\end{align}
\begin{align}
    &G(R,t>0) =-\frac{i}{\Omega} \sum_{k> k_\mathrm{F}} (D_1 +D_2 )
     e^{i\mathbf{k}\cdot\mathbf{R}},
\end{align}
where
\begin{align}
    C_1 &= Z e^{-iE_k t }\\
    C_2 &= (1-Z)e^{-i(E_k-\omega_\mathrm{p}) t }
\end{align}
\begin{align}
    D_1 &= Z e^{-iE_k t }\\
    D_2 &= (1-Z)e^{-i(E_k+\omega_\mathrm{p}) t }
\end{align}
where $E_k$ is the quasiparticle energy, $Z_k$ is the quasiparticle
renormalization factor, 
and $\omega_k$ is the plasmon energy. 
For simplicity, $Z_k$ and $\omega_k$ are 
assumed to be independent of $k$
and $E_k$ is taken to be a renormalized
free-electron gas dispersion:
\begin{equation}
    Z_k=Z,\qquad \omega_k=\omega_\mathrm{p},\qquad E_k=\alpha\varepsilon_k
    =\frac{\alpha}{2}k^2.
    \label{eq:Zk}
\end{equation}
For an electron gas of density $n_0$ the plasmon energy is given by
\begin{equation}
    \omega_\mathrm{p}=\sqrt{4\pi n_0}.
\end{equation}

For $t\neq 0$,
the exchange-correlation field can be obtained from the equation of motion:
\begin{align}
    V_\mathrm{xc}(R,t)& =\frac{1}{G(R,t)}\left[i\partial_t -h(R)\right] G(R,t).
    \label{eq:VxcR}
\end{align}
Since
\begin{equation}
 h(R) \exp{(i\mathbf{k}\cdot\mathbf{R})}=\frac{k^2}{2} \exp{(i\mathbf{k}\cdot\mathbf{R})},
\end{equation}
one finds for $t<0$
\begin{align}
  & \left[i\partial_t -h(R)\right] G(R,t<0)
  \nonumber\\
  &= \frac{i}{\Omega} \sum_{k\leq k_\mathrm{F}} (A_1+A_2)
    e^{i\mathbf{k}\cdot\mathbf{R}},
\end{align}
where
\begin{align}
    A_1 &= Z(E_k-\varepsilon_k)e^{-iE_k t } \\
    A_2 &= (1-Z)(E_k-\varepsilon_k-\omega_\mathrm{p})
    e^{-i(E_k-\omega_\mathrm{p}) t }
\end{align}
For $t>0$
\begin{align}
  & \left[i\partial_t -h(R)\right] G(R,t>0)
  \nonumber\\
  &=  - \frac{i}{\Omega}\sum_{k>k_\mathrm{F}} (B_1+B_2)
     e^{i\mathbf{k}\cdot\mathbf{R}},
\end{align}
where
\begin{align}
    B_1 &= Z(E_k-\varepsilon_k)e^{-iE_k t } \\
    B_2 &= (1-Z)(E_k-\varepsilon_k+\omega_\mathrm{p}) 
    e^{-i(E_k+\omega_\mathrm{p}) t }
\end{align}
\begin{figure}[t]
\begin{center} 
\includegraphics[scale=0.6, viewport=3cm 8cm 17cm 20cm, clip,width=\columnwidth]
{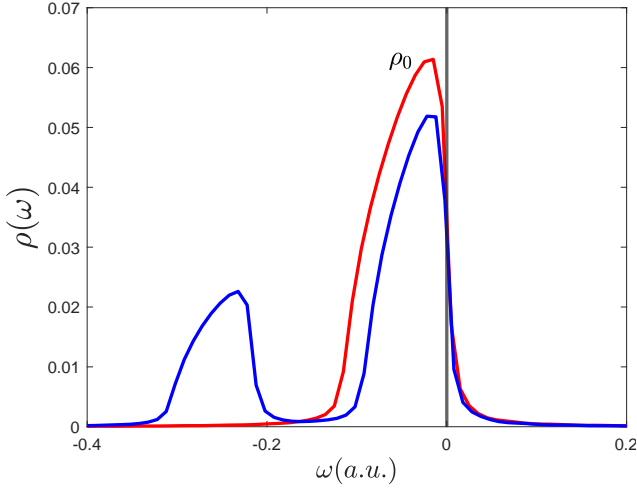}
\caption{
The hole spectral functions of the model interacting electron gas (blue) and
the non-interacting electron gas (red), labelled $\rho_0$.
The Fermi level is at the zero of the energy, indicated by a vertical line. 
The peak at around $\omega=-0.25$ is the plasmon satellite, located at one plasmon energy 
below the main quasiparticle peak. 
The model corresponds to $r_s=4$, giving a plasmon frequency $\omega_\mathrm{p}=0.217$.
A quasiparticle renormalization factor $Z=0.7$, a band-narrowing $\alpha=0.8$,
and a broadening $\eta=0.005$ have been used.
}
\label{fig:DOS}%
\end{center}
\end{figure}

\subsubsection{The exchange-correlation field and the kinetic potential}

Consider the case $t<0$.
Defining
\begin{align}
    I_n &= \int_0^{k_\mathrm{F}} dk k^{2n} e^{-i\alpha k^2 t/2}.
\end{align}
one obtains
\begin{align}
    \frac{1}{\Omega}\sum_{k\leq k_\mathrm{F}} A_1 &= \frac{1}{2\pi^2} Z\frac{\alpha-1}{2}I_2,\\
    \frac{1}{\Omega}\sum_{k\leq k_\mathrm{F}} A_2 & = \frac{1}{2\pi^2} (1-Z) e^{i\omega_\mathrm{p}t}
    \left(  \frac{\alpha-1}{2}I_2 - \omega_\mathrm{p} I_1 \right),
\end{align}
\begin{align}
    \frac{1}{\Omega}\sum_{k\leq k_\mathrm{F}} (C_1+C_2) 
    &= \frac{1}{2\pi^2}
    \left[ Z + (1-Z) e^{i\omega_\mathrm{p} t} \right] I_1.
\end{align}
Using the above results leads to
\begin{align}
  & \left[i\partial_t -h(R)\right] G(R,t<0) |_{R=0}
  = \frac{i}{2\pi^2} [a_2 I_2-a_1 I_1 ],
\end{align}
\begin{equation}
    G(0,t<0)=\frac{i}{2\pi^2} c_1 I_1
\end{equation}
where
\begin{align}
    a_1 &= \omega_\mathrm{p}(1-Z)e^{i\omega_\mathrm{p} t}
    \label{eq:a1}\\
     c_1 &= Z + (1-Z)e^{i\omega_\mathrm{p} t}.
    \label{eq:c1}\\
    a_2 &= \frac{\alpha-1}{2}c_1
    \label{eq:a2}\\ 
\end{align}

The exchange-correlation field becomes
\begin{equation}
    V_\mathrm{xc}(t<0)=\frac{1}{2}(\alpha-1)\frac{I_2}{I_1} -\frac{a_1}{c_1}.
\end{equation}

To calculate the difference in the kinetic potentials one needs
\begin{align}
    -i \nabla \cdot \mathbf{j} &=\frac{1}{\Omega}\sum_{k\leq k_\mathrm{F}} \frac{k^2}{2}(C_1+C_2)
    =\frac{1}{4\pi^2} c_1 I_2
    \\
    -i \nabla \cdot \mathbf{j}^\mathrm{KS}&=
    \frac{1}{\Omega}\sum_{k\leq k_\mathrm{F}} \frac{k^2}{2}e^{-itk^2/2}
    =\frac{1}{4\pi^2} I^0_2,
\end{align}
where%
\begin{align}
    I^0_n &= \int_0^{k_\mathrm{F}} dk k^{2n} e^{-i k^2 t/2}.
\end{align}
The temporal densities are given by
\begin{align}
    \rho_0(t<0)=-iG_0(0,t<0)= \frac{1}{2\pi^2} I^0_1,
\end{align}
\begin{align}
    \rho(t<0)=-iG(0,t<0)= \frac{1}{2\pi^2} c_1 I_1,
    \label{eq:rhot}
\end{align}
yielding
\begin{align}
    V_\mathrm{K}(t<0)&= \frac{1}{2}\frac{I_2}{I_1},
    \\
    V^\mathrm{KS}_\mathrm{K}(t<0)&= \frac{1}{2}\frac{I^0_2}{I^0_1}.
    \label{eq:VK}
\end{align}
It is interesting to note that the kinetic potential does not depend on the plasmon energy
and it cancels a term proportional to $I_2/I_1$ in the exchange-correlation field:
\begin{align}
    V_\mathrm{xc}+V_\mathrm{K}&=\frac{\alpha}{2}\frac{I_2}{I_1} -\frac{a_1}{c_1}.
    \label{eq:VKxc}
\end{align}


Using the relation
\begin{equation}
    \frac{\partial I_1}{\partial t}= -\frac{i\alpha}{2}I_2,
\end{equation}
the exchange-correlation field and the kinetic potential can be rewritten as
\begin{equation}
    V_\mathrm{xc}(t<0)=\frac{\alpha-1}{\alpha}
    i\partial_t \ln{I_1} -\frac{a_1}{c_1},
    \label{eq:Vxc}
\end{equation}
and
\begin{align}
    V_\mathrm{K}&= \frac{i}{\alpha}\partial_t\ln{I_1},
\end{align}
so that
\begin{align}
    V_\mathrm{xc}+V_\mathrm{K}&= i \partial_t\ln{I_1} -\frac{a_1}{c_1}.
\end{align}
The kinetic potential cancels a term in $V_\mathrm{xc}$ to give the correct band narrowing.

The first term of $V_\mathrm{xc}+V_\mathrm{K}$ when integrated over time from $0$ to $t$ is given by
\begin{align}
    -i\int_0^t dt' i
    \frac{\partial }{\partial t'} \ln{I_1}&=
    \ln{\frac{I_1(t)}{I_1(0)}}
    \label{eq:VxcQP}
\end{align}
and the second is given by
\begin{align}
    i\int_0^t dt' \frac{a_1}{c_1}&= i\omega_\mathrm{p}
    \int_0^t dt' \frac{(1-Z)e^{i\omega_\mathrm{p} t'}}{Z+(1-Z)e^{i\omega_\mathrm{p} t'}}
    \nonumber\\
    &= \ln{\left[ Z+(1-Z)e^{i\omega_\mathrm{p} t} \right]}.
    \label{eq:VxcSat}
\end{align}
Collecting the above two terms leads to
\begin{align}
    e^{-i\int_0^t dt' [V_\mathrm{xc}(t')+V_\mathrm{K}(t')]}=
    \left[ Z+(1-Z)e^{i\omega_\mathrm{p} t} \right] \frac{I_1(t)}{I_1(0)}.
\end{align}
Since $I_1(0)=2\pi^2\rho(r)$, the formal solution in Eq. (\ref{eq:formalsoln}) is then
\begin{align}
    \rho(r,t)=\frac{1}{2\pi^2} [Z+(1-Z)e^{i\omega_\mathrm{p} t}]  I_1(t),
\end{align}
which reproduces the temporal density
in Eq. (\ref{eq:rhot}).

\subsection{Results}

Atomic units are used throughout.
A quasiparticle renormalization factor $Z=0.7$, a band-narrowing parameter $\alpha=0.8$,
and a broadening $\eta=0.005$ have been used for all values of $r_s$.

In Fig. \ref{fig:DOS}, the hole spectral function of the model for $r_s=4$
is compared with that of the non-interacting
electron gas. The model essentially accounts for the quasiparticle band narrowing and the transfer of the 
quasiparticle weight to the plasmon satellite, located at one plasmon energy below the quasiparticle
band. For simplicity, only one plasmon is taken into account and there is no weight arising
from states above the Fermi level. The Fermi level of the interacting model has been adjusted to coincide
with that of the non-interacting one.

\begin{figure}[H]
\begin{center} 
\includegraphics[scale=0.6, viewport=3cm 4cm 18cm 23.5cm, clip,width=\columnwidth]
{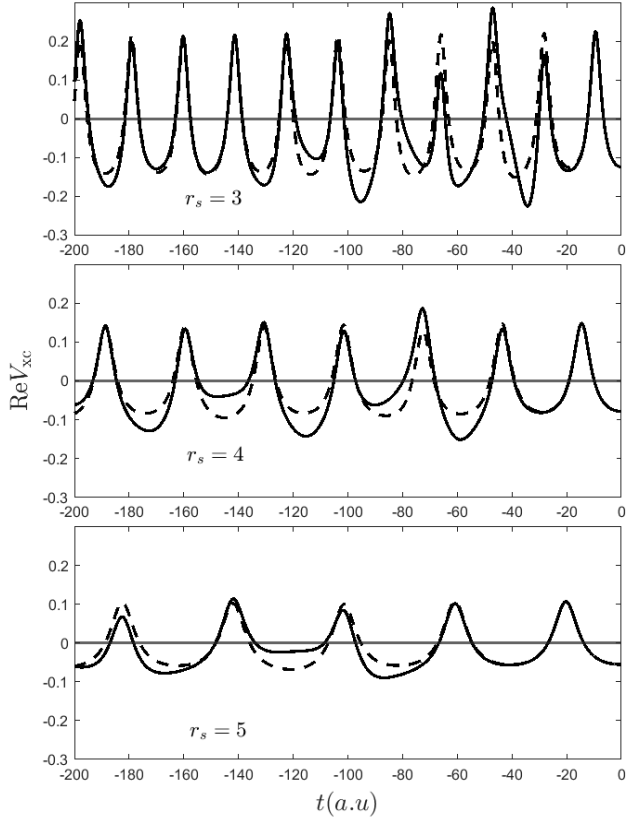}
\caption{
The real part of
the exchange-correlation potentials $V_\mathrm{xc}$ (dashed) and $\Tilde{V}_\mathrm{xc}$ (solid) as
defined in the text for $r_s=3,4,5$.
The difference, $\Delta V_\mathrm{K}=\Tilde{V}_\mathrm{xc}-V_\mathrm{xc}$, is shown in
Fig. \ref{fig:DVK}.
}
\label{fig:ReVxc}%
\end{center}
\end{figure}

\begin{figure}[H]
\begin{center} 
\includegraphics[scale=0.6, viewport=3cm 4cm 18cm 23.5cm, clip,width=\columnwidth]
{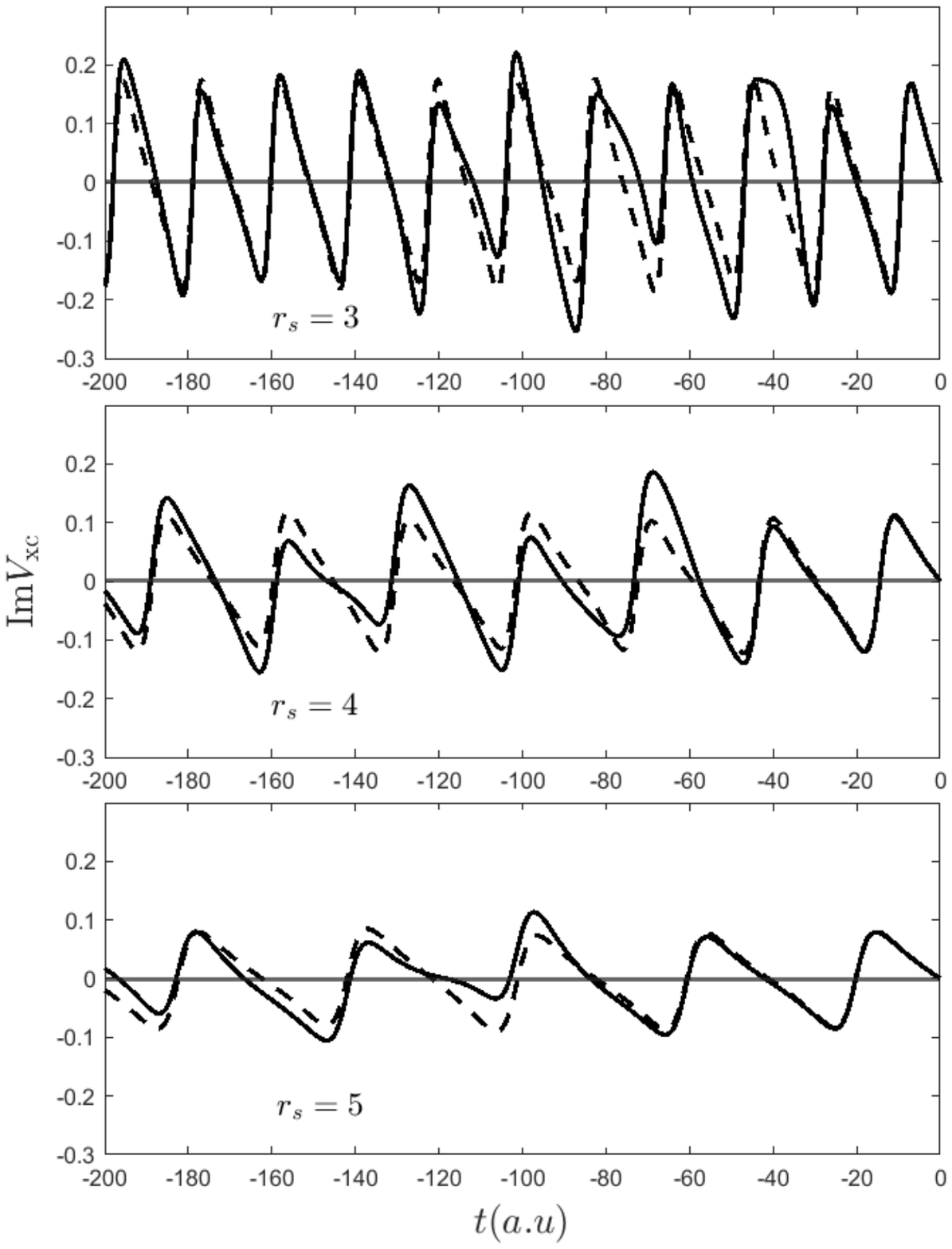}
\caption{
The imaginary part of
the exchange-correlation potentials $V_\mathrm{xc}$ (dashed) and $\Tilde{V}_\mathrm{xc}$ (solid) as
defined in the text for $r_s=3,4,5$.
The difference, $\Delta V_\mathrm{K}=\Tilde{V}_\mathrm{xc}-V_\mathrm{xc}$, is shown in
Fig. \ref{fig:DVK}.
}
\label{fig:ImVxc}%
\end{center}
\end{figure}

\begin{figure}[H]
\begin{center} 
\includegraphics[scale=0.6, viewport=3cm 4cm 18cm 24cm, clip,width=\columnwidth]
{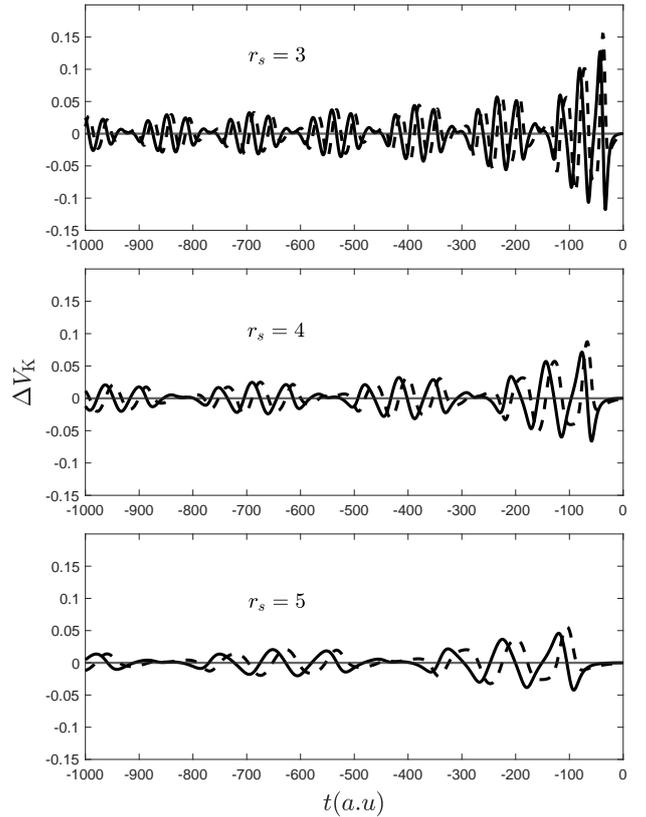}
\caption{
The real part (solid) and the imaginary part (dashed)
of the kinetic potential difference $\Delta V_\mathrm{K}=V_\mathrm{K} -V^\mathrm{KS}_\mathrm{K}$
for $r_s=3,4,5$.
}
\label{fig:DVK}%
\end{center}
\end{figure}

From the model Green function, the exchange-correlation fields can be extracted as detailed in the theory
section. The results are shown in Figs. \ref{fig:ReVxc} and \ref{fig:ImVxc} for the
real and imaginary parts of
$V_\mathrm{xc}$ and $\Tilde{V}_\mathrm{xc}$.
The former exhibits a more distinct periodicity whereas the latter appears to have a less well-defined 
periodicity. This can be understood from Fig. \ref{fig:DVK}, which shows
the difference between $\Tilde{V}_\mathrm{xc}$ and $V_\mathrm{xc}$. This difference, which is also the
difference in kinetic potential
between the interacting system and the non-interacting Kohn-Sham system, 
has a beat pattern which decreases in magnitude as $-t$ increases. The price of approximating the 
interacting kinetic potential by that of the Kohn-Sham system and transferring the difference into the
exchange-correlation field is a more irregular behavior of the latter.

\begin{figure}[H]
\begin{center} 
\includegraphics[scale=0.6, viewport=3cm 4cm 18cm 24cm, clip,width=\columnwidth]
{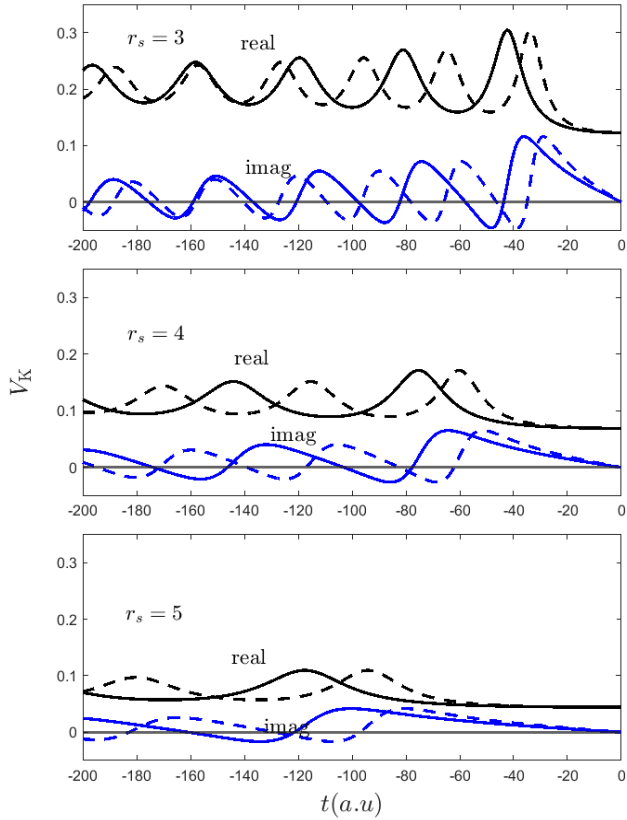}
\caption{
The real (black) and imaginary (blue) parts of the kinetic potentials $V_\mathrm{K}$ (solid)
and $V^\mathrm{KS}_\mathrm{K}$ (dashed) as defined in the text for $r_s=3,4,5$. 
}
\label{fig:VK}%
\end{center}
\end{figure}

In Fig. \ref{fig:VK} the kinetic potentials of the interacting system
and the non-interacting Kohn-Sham system are shown, both displaying well-defined oscillations.
The interacting kinetic potential mimics the behavior of
the Kohn-Sham kinetic potential but with a shifted phase, which appears to be time dependent. 
This suggests that rather than approximating
the interacting kinetic potential by that of the Kohn-Sham system and shifting the difference into 
the exchange-correlation field,
it could be more favorable to model directly the interacting kinetic potential by the Kohn-Sham one but
with a time-dependent shifted phase. 
The phase shift between the two kinetic potentials increases as the density is lowered, 
indicating that at high density the Kohn-Sham kinetic potential
better approximates the interacting kinetic potential. This is as anticipated as correlations
are expected to be less important as the density increases. 

There is a general trend of the exchange-correlation field and the kinetic potential as functions of $r_s$.
The smaller $r_s$ or the higher the density
the more oscillatory the quantities become. This is understandable since
the oscillatory behavior of the 
exchange-correlation field is determined to a large extent by the plasmon energy, which increases with 
the density.  The kinetic potential,
on the other hand, does not follow the same oscillatory behavior of the exchange-correlation field since
it does not depend explicitly on the plasmon energy, as can be seen in Eq. (\ref{eq:VK}). 
In the case of the kinetic potential, it is the Fermi wavevector that determines the oscillatory behavior,
which increases as the density increases or as $r_s$ decreases.


%


\section{Summary and conclusions}

The continuity equation for the temporal density has been derived, starting
from the recently proposed dynamical exchange-correlation field framework.
The current density, which is an unknown quantity in this equation, is approximated by
that of the Kohn-Sham system and the difference is transferred into the exchange-correlation field.
There remains 
the task of finding a good approximation for the exchange-correlation field, which should be
substantially simplified since
only the diagonal part is needed. 
If a good approximation for the exchange-correlation field can be constructed, the spectral function
can be readily calculated from an explicit
solution to the continuity equation. A model Green function of the interacting electron gas is used
to illustrate the key quantities in the proposed formulation. 

\begin{acknowledgments}
Financial support from the Knut and Alice Wallenberg (KAW) 
Foundation (Grant number 2017.0061)
and the Swedish Research Council (Vetenskapsrådet, VR, Grant number 2021\_04498) 
is gratefully acknowledged.
\end{acknowledgments}

\end{document}